# Sharp switching of the magnetization in Fe$_{1/4}$TaS$_2$

E. Morosan[1*], H. W. Zandbergen[2], Lu Li[3], Minhyea Lee[3],
J. G. Checkelsky[3], M. Heinrich[4], T. Siegrist[4], N. P. Ong[3] and R. J. Cava[1]

[1]Department of Chemistry, Princeton University, Princeton, NJ 08540, USA
[2]National Centre for HREM, Department of Nanoscience,
Delft Institute of Technology, Al Delft, The Netherlands
[3]Department of Physics, Princeton University, Princeton, NJ 08540, USA
[4]Bell Laboratories, Lucent Technologies, Murray Hill, NJ 07974, USA
*e-mail: emorosan@princeton.edu

Anisotropic magneto-transport measurements are reported on Fe$_{1/4}$TaS$_2$ single crystals grown by vapor transport. Both the magnetization and resistivity are extremely anisotropic, with the magnetic moments aligned parallel to the c crystallographic direction. Fe$_{1/4}$TaS$_2$ orders ferromagnetically below T$_C$ = 160 K and displays very sharp hysteresis loops in the ordered state for H∥c. The corresponding magnetoresistance is negative, and it qualitatively reproduces the features observed in the M(H) data, by showing a sharp drop around the critical field H$_s$ for the moment reversal. The magnetization switching time shows an unusual increase with increasing temperature. For field applied within the ab plane, the magnetization remains small and linear in field up to 5 T, and the magnetoresistance is positive and quadratic in field, with no visible hysteresis. The squareness of the H∥c M(H) loops and the high critical field for the magnetization switch (H$_s$ = 3.7 T at T = 2 K) allow us to classify Fe$_{1/4}$TaS$_2$ as a strong ferromagnet.

**Introduction**

Transition metal dichalcogenides, in their various polytypic forms, cover a wide spectrum of physical properties: insulators ($HfS_2$), semiconductors ($MoS_2$, $MX_2$ [M = Pt, Pd and X = S, Se]), semi-metals ($WTe_2$, $TcS_2$) or metals ($NbS_2$, $VSe_2$).[1] The strong anisotropy inherent to their low dimensionality often leads to charge density wave (CDW) transitions.[2] Superconductivity also appears, and competes with the CDW state in a numbers of these materials.[2-5] Intercalation of atoms and molecules between the $MX_2$ layers of the transition metal dichalcogenides leads to significant changes in their physical properties: the superconducting transition temperature changes as many $MS_2$ compounds are intercalated with organic molecules[6] or with Na or K atoms[7], or even if the intercalated compounds are further hydrated.[8] Cu intercalation of $TiSe_2$ leads to superconductivity.[9] Long range magnetic order occurs in many cases when $NbS_2$ and $TaS_2$ are intercalated with 3d-transition metals (T).[10-12] The structural and magnetic properties of $T_xMX_2$ depend strongly on the amount of intercalant x.[13-15]

In $Fe_xTaS_2$, ordered superlattices of the intercalated ions form for x = 1/4 or 1/3, giving rise to superstructures with a' = $2a_0$ and a' = $\sqrt{3}a_0$ respectively, where $a_0$ is the basic hexagonal lattice parameter of the $TaS_2$ array. The ferromagnetic ordering temperature ($T_C$) changes in a nonmonotonic manner for $0.20 \leq x \leq 0.34$.[13] The Weiss temperature decreases with increasing x, followed by a change of the magnetic interactions from ferromagnetic ($0.20 \leq x \leq 0.4$) to antiferromagnetic ($0.40 < x$).[15] $Fe_{1/3}TaS_2$ orders ferromagnetically around 35 K, and $T_C$ would be expected to decrease in the less concentrated $Fe_{1/4}TaS_2$ compound. However, the ferromagnetic ordering temperature in the latter is almost five times larger, around 160 K. No systematic study of the magnetic and transport properties has been performed on these compounds.

Here we report a detailed study of the anisotropic magneto-transport properties of $Fe_{1/4}TaS_2$ single crystals, with emphasis on their unusual field dependence; previously reported properties are elaborated. The magnetization and resistivity are extremely anisotropic, with the easy axis parallel to the c crystallographic direction (perpendicular to the $TaS_2$ planes). Below the ferromagnetic ordering temperature, the H∥c M(H) curves display a very sharp hysteresis loop, even at temperatures greater than half of $T_C$. The abrupt change in the direction of the moments is reflected in the transverse H∥c magnetoresistance data, which show a sharp drop at approximately the same field values as the magnetization data. In contrast, the H∥ab ρ(H) isotherms are quadratic in field and show no hysteresis.

**Experimental**

Single crystals of $Fe_{1/4}TaS_2$ were grown using iodine vapor transport. Fe, Ta and S powders in a 0.4: 1: 2 ratio were first sealed in evacuated silica tubes, slowly heated (~ 50°/ day) to 750°C and then annealed for 2 days at the high temperature. Subsequently, 20% $I_2$ (by mass) was added to the reacted powder, sealed in a 150 mm long silica tube with a 12 mm diameter. The tube was placed in a gradient furnace with the hot end temperature set at 1000°C and the cold end temperature kept at 900°C. After nine days, the hot end was cooled first to 800°C and subsequently kept ~100º lower than the cold end to minimize condensation of iodine vapor on the crystals formed towards the cold end. Next, the cold end temperature was decreased to 600°C, maintained constant for 2 days, and then the power on the hot end was turned off. Finally, the whole furnace was cooled to room temperature over a few hours. Large, thin hexagonal plates grew close to the cold end of the silica tube. Small amounts of iodine were noticeable on the surface, most of which evaporated when the crystals were warmed up above room temperature in an opened container. Characterization of the crystals by EDX and single crystal x-ray analysis showed that no iodine was incorporated in the lattice.

X-ray diffraction measurements were employed to characterize the samples. Room temperature data were recorded on a Bruker D8 diffractometer using Cu Kα radiation and a graphite diffracted beam monochromator (Bruker D8 Focus). Electron diffraction was performed at room temperature with a Philips CM200ST-FEG electron microscope, operated at 300 kV. Single crystal diffraction data were collected using an Oxford Diffraction Xcalibur 2 single crystal diffractometer equipped with a CCD and graphite monochromated MoKα radiation. The data were analyzed using the NRCVAX software.[16-18]

Magnetization measurements as a function of temperature and applied field M(H,T) were performed in a Quantum Design MPMS SQUID magnetometer (T = 1.8 - 350 K, H = 5.5 T). Anisotropic resistivity ρ(T,H) measurements with current parallel to the ab-plane were taken in a Quantum Design PPMS, using a standard four probe technique. The time-dependence of the switching field was measured using a thin copper wire pickup coil; the output voltage signal was amplified with PAR 113 pre-amplifier and then recorded with a digital oscilloscope.

**Results**

Single crystal X-ray diffraction measurements were performed on a crystal with dimensions 0.06×0.06×0.01 mm$^3$. The Fe$_{1/4}$TaS$_2$ data were consistent with the reported 2H-TaS$_2$ type structure.[15] In the 2H polymorphic form of TaS$_2$, the Ta atoms are in a trigonal prismatic coordination with the chalcogens. The single crystal X-ray structure analysis on Fe$_{1/4}$TaS$_2$ showed that this basic crystal structure is preserved, with the Fe ions intercalated in an ordered fashion in the octahedral sites between the TaS$_2$ layers (Fig. 1a). Fe$_{1/4}$TaS$_2$ is thus isostructural to MnTa$_4$S$_8$.[19] The refined structural parameters are presented in Table I.[1] Electron diffraction on Fe$_{1/4}$TaS$_2$ shows the presence of reflections corresponding to the basic trigonal structure and also the 2a superstructure reflections (Fig. 1b). This observation, together with the single crystal structure refinement, reflects that the Fe sublattice is ordered in the layer between the TaS$_2$ planes. The electron diffraction patterns with the c* axis in the diffraction plane (Fig. 1c) showed many crystallites to have strong streaking of the superreflections along c*, indicating that the Fe is well ordered in plane, but that in some cases stacking of the planes along c is correlated only over finite distances. Crystallites with superreflections with no streaking were also observed, in which case the superstructure was 2a, c. The crystal used for single crystal X-ray structural refinement showed an ordered superstructure.

Figure 2 shows the anisotropic inverse susceptibility data of Fe$_{1/4}$TaS$_2$, in an applied field H = 0.01 T, after the diamagnetic contributions $\chi_0$ have been subtracted for each field orientation. The $\chi_0$ values, determined by fitting the direct susceptibility data (inset, Fig. 2) to a Curie-Weiss law $\chi(T) = \chi_0 + C/(T - \theta_C)$ above ~160 K, are $\chi_0 = -5*10^{-4}$ emu/(mol Fe) and $-4*10^{-3}$ emu/(mol Fe) for H||c and H||ab, respectively. The high-temperature inverse susceptibilities are linear in temperature, confirming the Curie-Weiss behavior. When the field is applied perpendicular to the hexagonal plates (H||c), a linear fit of the $(\chi - \chi_0)^{-1}$ data (full squares, Fig. 2) for T > 200 K yields a Weiss temperature $\theta_c$ = 162 K, and an effective moment $\mu^c_{eff}$ = 5.03 $\mu_B$. For the other field orientation (H||ab, full triangles, Fig. 2) the analogous parameters are $\theta_{ab}$ = 108 K and $\mu^{ab}_{eff}$ = 5.12 $\mu_B$. After the subtraction of the diamagnetic contribution for each direction, the anisotropic magnetic susceptibilities $\Delta\chi = \chi - \chi_0$ can be used to estimate the average susceptibility $\chi_{ave} = (2\Delta\chi_{ab} + \Delta\chi_c)/3$. As expected, the inverse average susceptibility $\chi^{-1}_{ave}$ (open circles, Fig. 2) is linear at high temperatures, and a linear fit yields a Weiss temperature $\theta_{ave} = 135$ K and an effective moment $\mu^{ave}_{eff} = 5.01$ $\mu_B$. All susceptibility data is consistent with high-spin $Fe^{2+}$ ($^5D_4$) configuration, for which an effective moment $\mu_{eff} = 4.90$ $\mu_B$ is expected. The inverse susceptibility curves allow an estimate of the anisotropy of the paramagnetic susceptibility as $\chi_c / \chi_{ab} \approx 6$ at T = 170 K, a temperature just above where magnetic ordering is apparent.

---

[1] Further details of the crystal structure investigation may be obtained from Fachinformationszentrum Karlsruhe, 76344 Eggenstein-Leopoldshafen, Germany (fax: (+49)7247-808-666; e-mail: crysdata@fiz-karlsruhe.de, http://www.fiz-karlsruhe.de/ecid/Internet/en/DB/icsd/depot_anforderung.html) on quoting the deposition numberCSD-417150.

The temperature dependent magnetization of $Fe_{1/4}TaS_2$ in the ordered state is also highly anisotropic, as evidenced by comparing the H||c data (Fig. 3a) with similar measurements for H||ab, displayed in Fig. 3b. When the temperature is lowered, ferromagnetic ordering occurs around $T_C = 160$ K. For the lowest measured magnetic field (H = 0.01 T) a sharp increase of the H||c magnetization is observed just below $T_C$ (Fig. 3a); rounding of the transition near the ordering temperature is observed upon increasing the applied field, as expected for ferromagnets due to spin fluctuations close to $T_C$. Cooling below 155 K results in different magnetization behavior depending on whether the data were collected upon zero-field cooling (ZFC, full symbols in Fig. 3a) or field-cooling (FC, open symbols in Fig. 3a). The FC curve increases and plateaus around $22*10^3$ emu mol$^{-1}$ Fe $\approx 3.9$ $\mu_B$ per Fe for H ≥ 0.5 T. This magnetization value corresponds to the saturated moment $\mu_{sat} = 4.0$ $\mu_B$ expected for $Fe^{2+}$ at high fields and low temperatures. Conversely, for the lowest applied field H = 0.01 T, the magnetization is not fully saturated, resulting in a much smaller FC value at low temperatures, as can be seen in Fig. 3a. The ZFC magnetization has a local maximum at the irreversibility temperature $T_{irr} = 155$ K, which displays a strong field dependence and is reduced to about 7 K at our maximum applied field (H = 5.5 T). For H||ab, the magnetization behavior is similar to that for H||c; the most notable differences are the much smaller overall magnetization values for field applied within the ab plane, and also the much weaker field dependence ($T_{irr} \approx 140$ K for H = 1.5 T) of the irreversibility temperature, $T_{irr}$. This is expected for the "hard" direction of the applied field in a strong ferromagnet. Also, a small kink is noticeable at 40 K (Fig. 3b, inset), and it may be intrinsic to the spin system of $Fe_{1/4}TaS_2$ or due to the presence of a very small amount of $Fe_xTaS_2$, x ≠ ¼ in the crystal used for this measurement.

A remarkable behavior of the magnetization is observed in measurements of M(H). At T = 2 K (Fig. 4) the strong anisotropy inferred from the magnetic susceptibility is confirmed. The H||ab magnetization is very low ($M_{ab}[5.5T] < 0.25$ $\mu_B$ per Fe) and linear up to the maximum applied field. The H||c magnetization saturates around 4 $\mu_B$ per Fe, displaying a very sharp hysteresis loop ($\Delta H < 0.01$ T, where $\Delta H$ is the measured width of the transition). This is the manifestation of a very strong ferromagnet with a threshold field $H_s \approx 3.7$ T and a nearly field-independent magnetization of $\pm$ 4 $\mu_B$, the full saturated moment of $Fe^{2+}$. As figure 5 shows, the hysteresis loops remain sharp throughout the ordered state (T < 160 K). This suggests that the crystal may be a nearly single-domain ferromagnet for H < $H_s$ and H > $H_s$, and that rapid switching of the orientation of the spins occurs at the threshold field $H_s$. The very small difference in the T = 2 K saturation magnetization after the first hysteresis cycle could be a consequence of a small initial misalignment of the sample, corrected by the magnetic torque which orients the thin plate exactly perpendicular to the field, as H is cycled between – 5.5 T and 5.5 T.

The spontaneous magnetization, $M_s$, determined by extrapolating the high-field magnetization to H = 0, and the critical field values, taken as the H > 0 field values where the M(H) curves intercept the M = 0 axis, are shown in Fig. 6 as a function of temperature. The reduced spontaneous magnetization $m_s(T) = M_s(T)/M_s(0)$ (symbols in Fig. 6a) is best described by

$$m_s(t) = [1 - s\, t^{3/2} - (1 - s)\, t^{5/2}]^{\beta}, \text{ where } t = T/T_c.$$

The above equation was proposed[20-21] such that, at low temperatures (t → 0), it obeys Bloch's $T^{3/2}$ power law, and, around the Curie temperature (t → 1), it describes the expected critical behavior of the Heisenberg model[22] $m_s(T) = [1 - t]^{1/3}$ when $\beta = 1/3$. The solid line in Fig. 6a corresponds to the fit using the above equation with s = 0.26 and $\beta \approx 0.349$. The critical exponent, $\beta$, determined from our data is intermediate between the value expected for metallic ferromagnets, $\beta = 1/3$, and the 3D Heisenberg value often observed for magnetic insulators, $\beta = 0.369$.[23] The threshold field $H_s$ (Fig. 6b) decreases with increasing temperature from $H_s(2K) \approx$ 3.7 T to less than 0.05 T, just below $T_C$. When plotted as $\ln(H_s)$ vs. T, an approximately linear behavior is observed at low temperatures, yielding a "Debye-like" temperature dependence $H_s(T) \propto e^{-aT}$ over much of the temperature range.

The observed sharp M(H) loops suggest that the switching of the magnetic moment at $H_s$ may be quite rapid. Depending on sample size and domain structure, switching times in ferromagnets have been observed to range from a few nanoseconds in ferromagnetic films[24] to a few milliseconds in hard bulk magnets[25] and usually increase slowly on cooling. The switching time is expected to be dependent on how fast the extra energy created by the moment-flipping gets damped by the eddy currents and can be converted to heat.[26,27]

Here we measure the switching time of $Fe_{1/4}TaS_2$ by recording the electromotive force (EMF) generated on a pickup coil around the crystal during the switching of the magnetic moments' orientation. As the applied magnetic field is swept slowly and passes the threshold field $H_s$, the sudden reversal of magnetic moment leads to a fast change of magnetic flux inside the pickup coil; in turn, this induces a pulse of EMF voltage due to the Faraday Effect. We measure the EMF pulse around the positive and negative switching field $H_s$, by sweeping the field both up and down. Fig. 7a displays a typical pulse signal at T = 4.4 K. For $H \rightarrow H_s$ (full symbols, Fig. 7a) the flux changing rate first accelerates, reaches the terminal rate, fluctuates and finally falls off. The fine structure at the top of the pulse suggests that, on the time scale of these measurements, the domain wall motion is not uniform. However, this behavior is not accidental, as is indicated by the fact that the pulse for the opposite field orientation (open symbols, $H \rightarrow -H_s$) shows the same features. It may be that the macroscopic crystal consists of multiple domains in which the domain wall motion occurs at slightly different speeds. A distribution of defects within a single domain can also cause the domain wall speed to be different in different regions of the macroscopic crystal, causing the kinds of features observed.

We define the pulse width, τ, as the interval between the moments when the pulse signal becomes less than 10 % of the maximum at each temperature, as indicated in Fig. 7a, and take that as the switching time for the bulk sample. Measurements at temperatures between 4.4 K and 18 K allow us to study the temperature dependence of the pulse width. Fig. 7b shows the pulse signal at various representative temperatures. The resulting temperature dependence of the switching time is plotted in Fig. 8. Surprisingly, the domain switching time decreases on cooling by a factor of approximately 10 as the temperature decreases from 18 K to 4 K. The behavior is approximately linear in this temperature range. As shown below, the resistivity ρ(T) is virtually constant in this temperature range, suggesting that the eddy currents cannot be the major factor in determining the switching time.

The zero-field temperature dependence of the resistivity in the basal plane of $Fe_{1/4}TaS_2$ is shown in Fig. 9. At room temperature, the resistivity is metallic, around 50 μΩ cm. As the temperature is decreased, the resistivity shows weak linear temperature dependence above $T_C$.

This is followed by a drastic decrease of the resistivity in the ferromagnetic state, due to loss of spin-disorder scattering. This behavior is very similar to what has been previously reported for $Fe_xTaS_2$ x = 0.28[13], a composition close to our x = ¼.

Anisotropic magneto-transport measurements were also performed. Due to the thin plate geometry of the crystals, these measurements were restricted to transverse magnetoresistance, with current flowing within the ab-plane and the field perpendicular to the current, H||ab or H||c. Quadratic field-dependence of the low-temperature transverse magnetoresistance with both applied field and current in the basal plane (H || ab, $i \perp H$) is observed (Fig. 10a) up to H = 5 T, with no apparent irreversibility. However, when field is applied in the other orientation (H || c, $i$ || ab), the magnetoresistance qualitatively reproduces the features observed in the M(H) data (Fig. 10b). As the magnetic field is increased from 0 up to ~ 2 T at T = 6 K, the magnetoresistance (full symbols) remains nearly unchanged, corresponding to the zero-magnetization region on the virgin M(H) curve (open symbols). Increasing the field beyond 2 T results in a rapid decrease of the magnetoresistance, consistent with the alignment of the magnetic domains and reflected in the corresponding region of the magnetization curve. It is surprising that even after the magnetization reaches its saturation value around H = 3 T, the magnetoresistance continues to decrease almost linearly with field up to the maximum applied field of 5 T, though at a slower rate than before the domain reorientation. When the field is next decreased, ρ(H) follows the same nearly linear dependence; moreover, this trend continues when the field direction is reversed, and is followed by a sharp drop in resistivity around the same negative field value where the magnetization switches direction. Further varying the field between -5 T and 5 T yields an almost symmetric ρ(H) curve, in good agreement with the magnetization loop at the same temperature (T = 6 K). Minor misalignment of the current contacts or small torque on the sample during the field sweep could account for the slight ρ(H) asymmetry.

The qualitative field-dependence of ρ reflects the properties of the magnetic state inferred from the M(T,H) data (*i.e.,* the magnetic moments order ferromagnetically along the c-axis). Consequently, the spin-scattering increases for fields applied along the "hard" direction (H||ab), leading to a magnetoresistance quadratically increasing with H (Fig. 10a). For field applied in the "easy" direction (H||c), increasing applied magnetic fields align the magnetic moments in the ferromagnetic state; this results in loss of spin disorder scattering, and thus a decreasing (negative) magnetoresistance (full symbols, Fig. 10b). A sudden enhanced magnetoresistance occurs when the magnetization changes direction at $H_s$, and this can be attributed to the domain reorientation parallel to the direction of the field. The cause of the decrease in magnetoresistance subsequent to the domain reorientation remains an open question, particularly since $M(H>H_s)$ is virtually constant.

**Discussions and conclusions**

$Fe_{1/4}TaS_2$ is a member of the series of intercalated layered dichalcogenides $M_xTaS_2$. Various amounts of transition metal can be intercalated between the $TaS_2$ layers, yielding either ferromagnetic or antiferromagnetic ordering in the respective compounds. The x = 1/4 and 1/3 compositions are the two ordered structures in this class of materials, and $Fe_{1/3}TaS_2$ shows ferromagnetic ordering at a $T_C$ much smaller than the less concentrated $Fe_{1/4}TaS_2$ compound. A detailed characterization of the anisotropic magnetic and transport properties of the latter compound reveals even more intriguing properties: a collective reorientation of the magnetic moments occurs at a high field value (H = 3.7 T at T = 2 K), allowing the classification of $Fe_{1/4}TaS_2$ as a strong ferromagnet. (Demagnetizing effects can be considerable for H||c in the thin-plate geometry of these crystals. We estimate the demagnetizing field to be around 0.7 T. Consequently the effective field at which the switching occurs is still quite high, around 3.0 T.) The squareness of the hysteresis loop, calculated as the ratio of the applied field where M is reduced to 10% of M(H=0), to the threshold field $H_s$, is larger than 0.9 throughout the ferromagnetic state. This is comparable to the squareness of the M(H) loops of strong permanent magnets (*e.g.*, $SmCo_5$[28] or Nd-Fe-B[29]), based on the shape of the magnetization curves. A sharp magnetization switch is observed in a single particle of $SmCo_5$ (diameter ~ 5 μm) subjected to thermal and chemical treatments. In $Fe_{1/4}TaS_2$ sharp magnetization reorientation (ΔH < 0.01 T) is observed in as-grown macroscopic crystals and is independent of the sample size. Neither optimal $SmCo_5$ nor Nd-Fe-B shows a collective switch of the magnetization orientation like $Fe_{1/4}TaS_2$, as their M(H) loops are more rounded, or display two or more steps. We surmise that the crystallographic short range order along c in $Fe_{1/4}TaS_2$ might facilitate the rapid switch of the magnetization direction at $H_s$. $Fe_{1/4}TaS_2$ has a Curie temperature $T_C$ of 160 K, much smaller than that of $SmCo_5$ ($T_C \approx 950$ K) or Nd-Fe-B ($T_C \approx 600$ K) but still fairly accessible. Although the energy product $(BH)_{max}$ for $Fe_{1/4}TaS_2$ is less than 1 MGOe, 20 to 50 times smaller than that of the well-known permanent magnets, the high threshold field of the Fe-intercalated dichalcogenide material renders it as a stable, strong ferromagnet and a potential candidate for applications at low temperatures.

A question that remains unelucidated with regards to the sharp magnetization switch in $Fe_{1/4}TaS_2$ concerns the temperature dependence of the switching field $H_s$ shown in Fig. 6b. Additional studies of the time-dependence of the switching field $H_s$ at low temperatures, together with Hall effect and angular dependent magnetoresistance measurements[30] are currently underway; these are expected to shed more light on the unusual magnetic behavior observed in $Fe_{1/4}TaS_2$. Finally, it would be interesting to clarify whether these unusual magnetic properties are indeed correlated with the crystallographic disorder along the c axis. To this end, measurements of the crystal structure and physical properties of $Fe_{1/4}TaS_2$ with structural long range order along c, if possible, would be desirable.

**Acknowledgements**

This research was supported primarily by the US DOE-BES solid state chemistry program, and, in part, by the US NSF MRSEC program grant DMR 0213706.
.

**Table I.** Crystal structure parameters for $Fe_{1/4}TaS_2$, space group *P 63/m m c*, a = 6.6141(15) Å, c = 12.154(3) Å (No. 194), Z = 2, $R_F$ = 0.037, $R_w$ = 0.043 (773 independent reflections, 411 reflections used [I > 2.5 σ(I)], 19 parameters).

| Atom  | x           | y       | z          | $B_{iso}$  |
|-------|-------------|---------|------------|------------|
| Ta (1)| 0.49507(6)  | 0.50493 | 3/4        | 0.520(16)  |
| Ta(2) | 0           | 0       | 1/4        | 0.511(19)  |
| Fe    | 0           | 0       | 0          | 1.07(8)    |
| S (1) | 2/3         | 1/3     | 0.1191(4)  | 0.64(8)    |
| S (2) | 0.83177(25) | 0.16823 | 0.62233(2) | 0.57(6)    |

**References**

[1] Wilson, J. A. & Yoffe, A. D., *Adv. Phys.* **28**, 193 - 335 (1969).

[2] Wilson, J. A., Di Salvo, F. J. & Mahajan, S., *Adv. Phys.* **24**, 117 - 201 (1975).

[3] Nagata, S., Aochi,T, Abe, T, Ebisu, S., Hagino, T., Seki, Y., Tsutsumi, K., *J. Phys. Chem. Solids* **53**, 1259 - 1263 (1992).

[4] Kumakura, T., Tan, H., Handa, T., Morishita, M. & Fukuyama, H., *Czech. J. Phys.* **46**, 2611 - 2612 (1996).

[5] Morris, R. C., Coleman, R. V., Bhandari, R., *Phys. Rev. B* **5**, 895 (1972).

[6] Gamble, F. R., Osiecki, J. H., Cais, M., Pisharody, R., DiSalvo, F. J., Geballe, T. H., *Science* **174**, 493 (1971).

[7] Somoano, R. B., Rembaum, A., *Phys. Rev. Lett.* **27**, 402 (1971).

[8] Johnston, D. C., *Mater. Res. Bull.* **17**, 13 (1982).

[9] Morosan, E., Zandbergen, H. W., Dennis, B. S., Bos, J. W. G., Onose, Y., Klimczuk, T., Ramirez, A. P., Ong, N. P., Cava, R. J., *Nature Physics* **2**, 544 (2006)..

[10] Friend, R. H., Beal, A. R., Yoffe, A. D., *Phil. Mag. B* **35**, 1269 (1977).

[11] Parkin, S. S. P., Friend, R. H., *Phil. Mag. B* **41**, 65 (1980).

[12] Parkin, S. S. P., Friend, R. H., *Phil. Mag. B* **41**, 95 (1980).

[13] Eibshutz, M., Mahajan, S., DiSalvo, F. J., Hull, G. W., Waszczak, J. V., *J. Appl. Phys.* **52**, 2098 (1981).

[14] Dijkstra, J., Zijlema, P. J., Bruggen, C. F. Van, Haas, C., deGroot, R. A., *J. Phys.: Condens. Matter* **1**, 6363 (1989).

[15] Narita, H., Ikuta, H., Hinode, H., Uchida, T., Ohtani, T., Wakihara, M., *J. Solid State Chem.* **108**, 148 (1994).

[16] Gabe, E. J., Le Page, Y., Charland, J.-P., Lee, F. L., White, P. S. *J. Appl. Cryst.* **22**, 384 (1989).

[17] "International Tables for X-Ray Crystallography", vol. IV (1974), Kynoch Press, Birmingham, England.

[18] Larson, A. C. "Crystallographic Computing" (1970), Munksgaard, Copenhagen.

[19] van Laar, B., Rietveld, H. M., Ijdo, D. J. W., *J. Solid State Chem*. **3**, 154 (1971).

[20] Kuz'min, M. D., *Phys. Rev. Lett*. **94**, 107204 (2005).

[21] Kuz'min, M. D., Richter, M., Yaresko, A. N., *Phys. Rev. B* **73**, 100401(R) (2006).

[22] Callen, E., Callen, H. B., *J. Appl. Phys* **36**, 1140 (1965).

[23] Kuz'min, M. D., Tishin, A. M., *Phys. Rev. Lett*. A **341**, 240 (2005).

[24] Smith, D. O. *J. Applied Phys*. **29**, 264 (2006).

[25] Galt, J. K. *Phys. Rev*. **85**, 664 (1952).

[26] Kittel, C., Galt, J. K. *Solid State Phys*. **3**, 437 (1956).

[27] Soohoo, R. F. *Magnetic Thin Films* (1965).

[28] Zijlstra, H., *J. Applied Phys*. **42**, 1510 (1971).

[29] Kawai, T., Ma, B. M., Sankar, S. G., Wallace, W. E., *J. Applied Phys*. **67**, 4610 (1990).

[30] Lee, Minhyea, Li, Lu, Checkelsky, J. G. (unpublished).

**Fig.1.** (a) The crystal structure of $Fe_{1/4}TaS_2$. Electron diffraction patterns of the $Fe_{1/4}TaS_2$ reciprocal lattice (a) along the [001] and (b) perpendicular to the [001] direction. The basic trigonal structure and the 2a superstructure reflections seen in (b), with the supercell outlined and the unit cell indicated; in (c) strong streaking of the superreflections along c* can be seen.

**Fig.2.** Inverse anisotropic susceptibilities at H = 0.01 T (full symbols), after subtraction of the temperature independent contribution $\chi_0$ in each measured direction. Open symbols represent the inverse average susceptibility $1/\chi_{ave}$, where $\chi_{ave} = (2\Delta\chi_{ab} + \Delta\chi_c)/3$ and $\Delta\chi = \chi - \chi_0$. Solid lines are linear fits to the inverse Curie-Weiss law. Inset: anisotropic H = 0.01 T susceptibility data.

**Fig.3. (color on-line)** Anisotropic temperature-dependent magnetization data M(T), zero-field cooled (full symbols) and field-cooled (open symbols) for (a) H||c and (b) H||ab and various applied fields.

**Fig. 4.** T = 2 K M(H) curves for H||c (triangles) and H||ab (crosses).

**Fig. 5. (color on-line)** H||c M(H) loops at T = 2, 4, 6, 8, 10, 20, 40, 60, 80, 100, 150 and 200 K.

**Fig. 6.** (a) Reduced spontaneous magnetization $m = M_s(T)/M_s(0)$ vs. $t = T/T_C$ (full symbols) and the fit to $m(t) = [1 - s\, t^{3/2} - (1-s)\, t^{5/2}]^\beta$ (solid line). (b) H||c $H_s - T$ phase diagram on a semi-log scale, where $H_s$ represents the magnetization switching field, with the dotted line representing a linear fit at low temperatures.

**Fig 7. (color on-line)** (a) Example of a EMF pulse signal proportional to $d\Phi/dt$ for T = 4.4 K. $\Phi$ is the magnetic flux inside the coil and is given by $\Phi = \mu_0\,(H + M)\cdot nS$, where $S$ in the cross session of the coil and n is the number of turns. The upper (lower) curve - closed (open) symbols - corresponds to measurements around $H_s > 0$ (< 0). (b) EMF signal of $Fe_{1/4}TaS_2$ at T = 4.4 K, 5.0 K, 5.7 K, 6.0 K, 8.0 K, 14.0 K and 18.0 K. The magnetic field is swept at a rate of 200 mT/min and the gain of the pre-amplifier is set to 200 – 2000; the EMF voltage is normalized to the corresponding gain.

**Fig 8.** The temperature dependence of the pulse width $\tau$ (symbols), with the dotted line emphasizing the almost linear increase of $\tau$ with temperature.

**Fig. 9.** Zero-field temperature-dependent resistivity of $Fe_{1/4}TaS_2$ for $i\,||\,ab$.

**Fig. 10. (color on-line)** (a) Transverse magnetoresistance $\Delta\rho/\rho_0$ (symbols) for H||ab and $i \perp$ H, with the dotted line emphasizing the quadratic field behavior of $\rho_{ab}$. (b) T = 6 K H||c magnetization (open symbols) and magnetoresistance ($i \perp$ H, full symbols) loops.

Fig. 1.

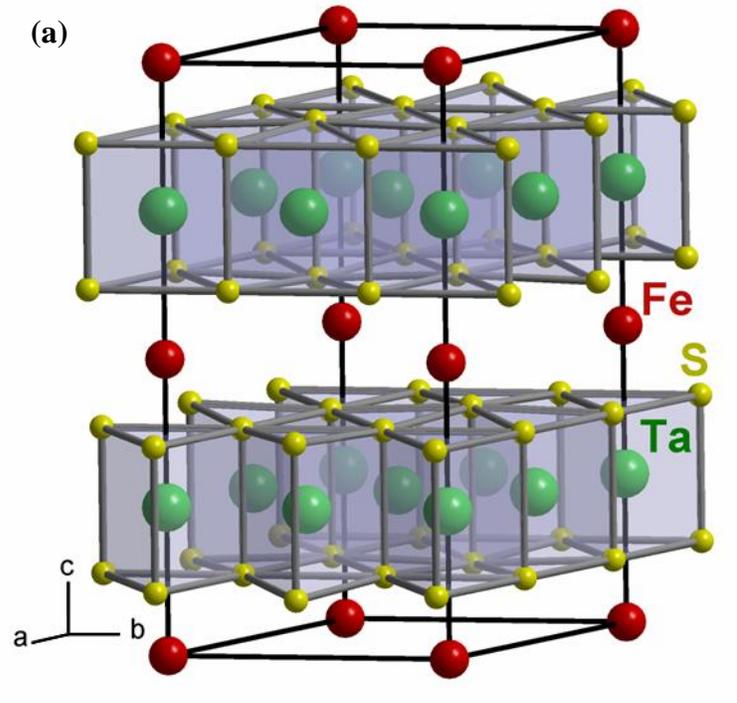

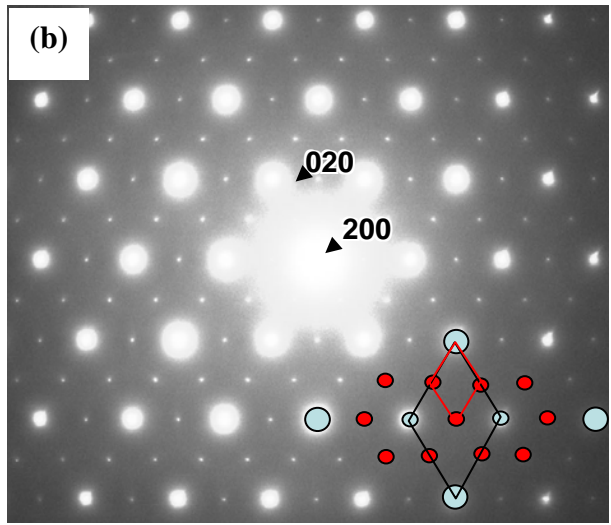
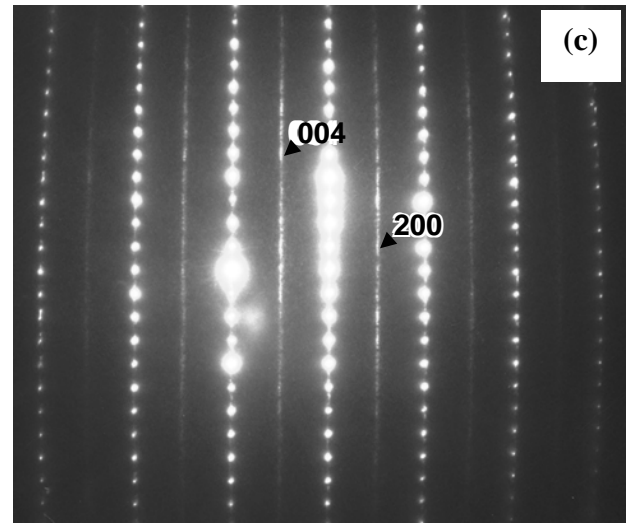

Fig.2.

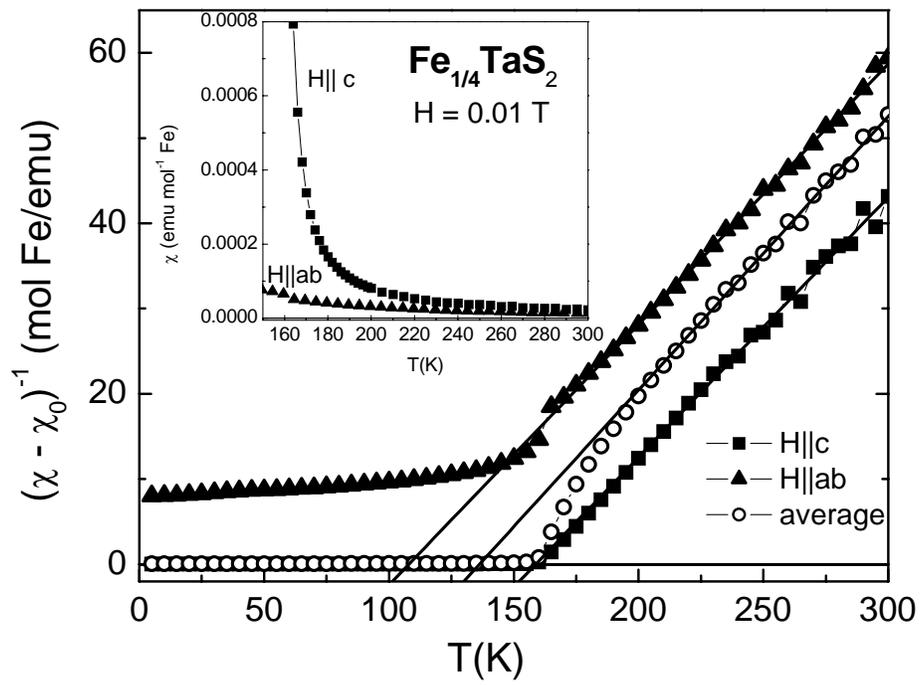

Fig. 3.

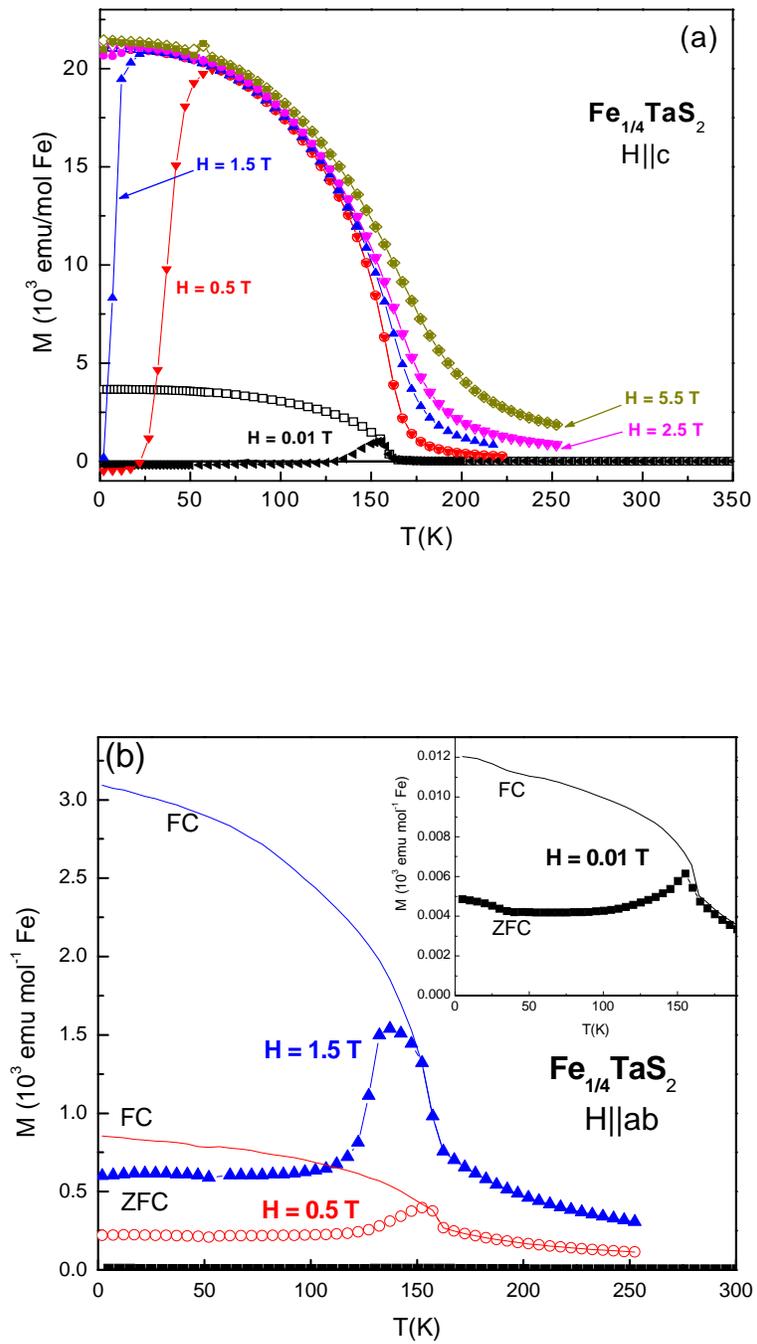

Fig. 4

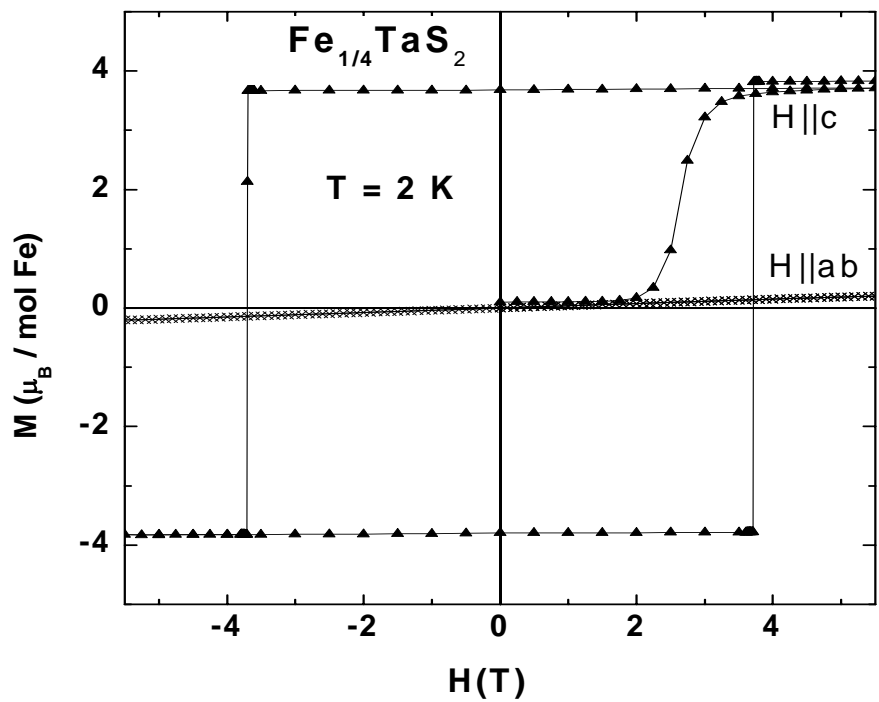

Fig.5

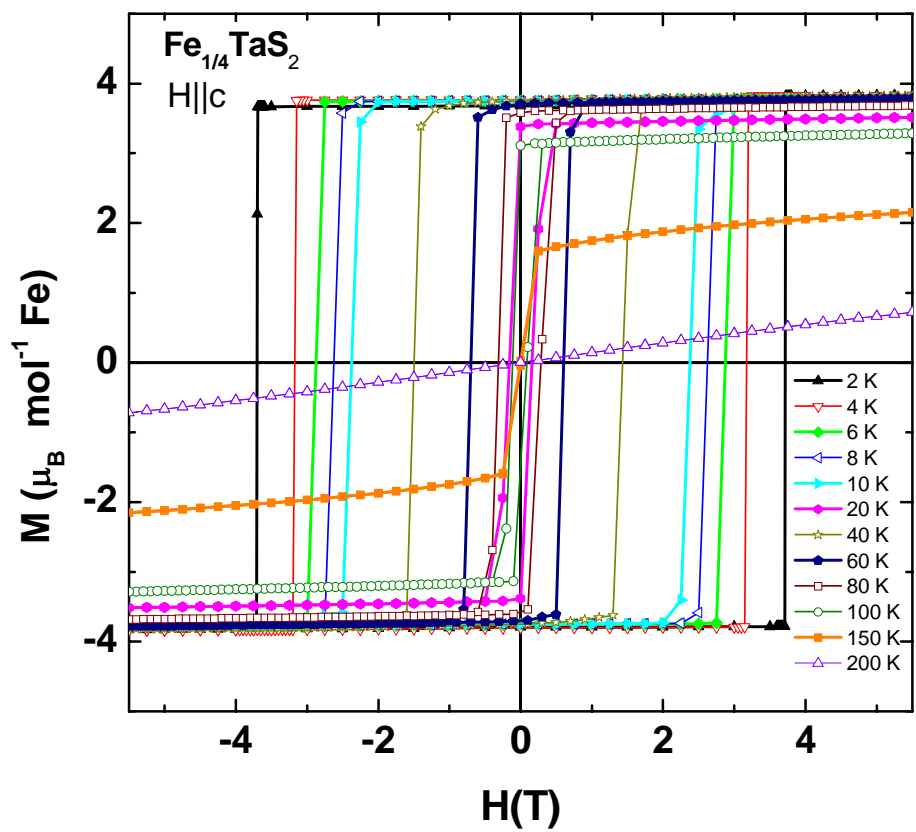

Fig.6

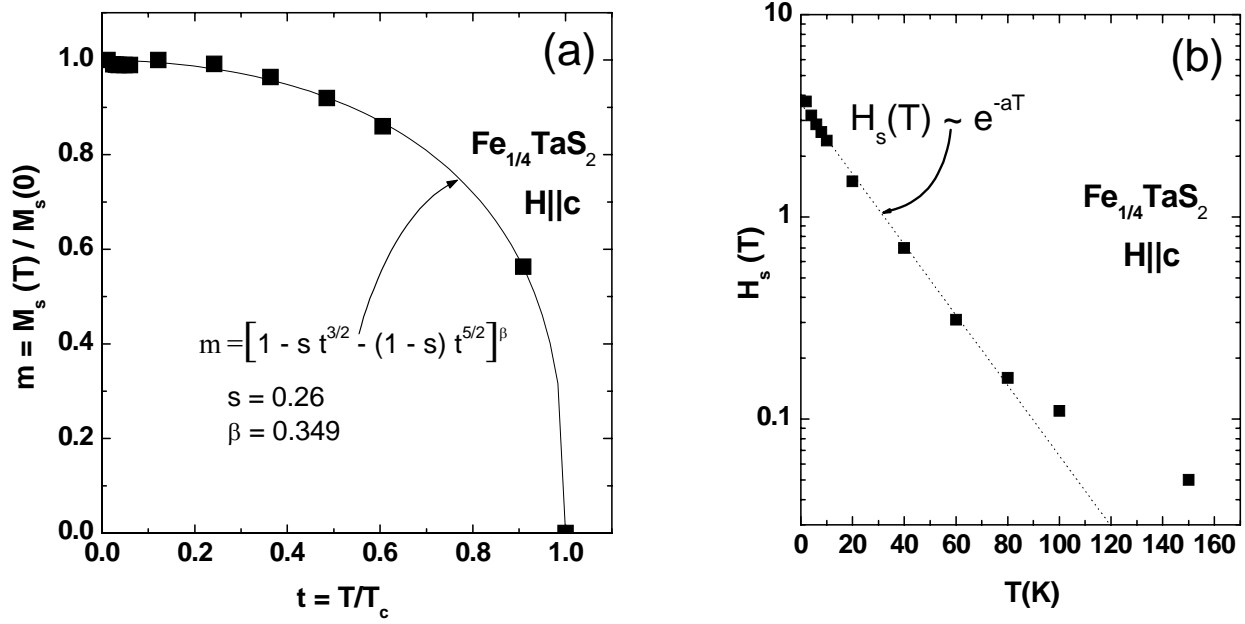

Fig.7

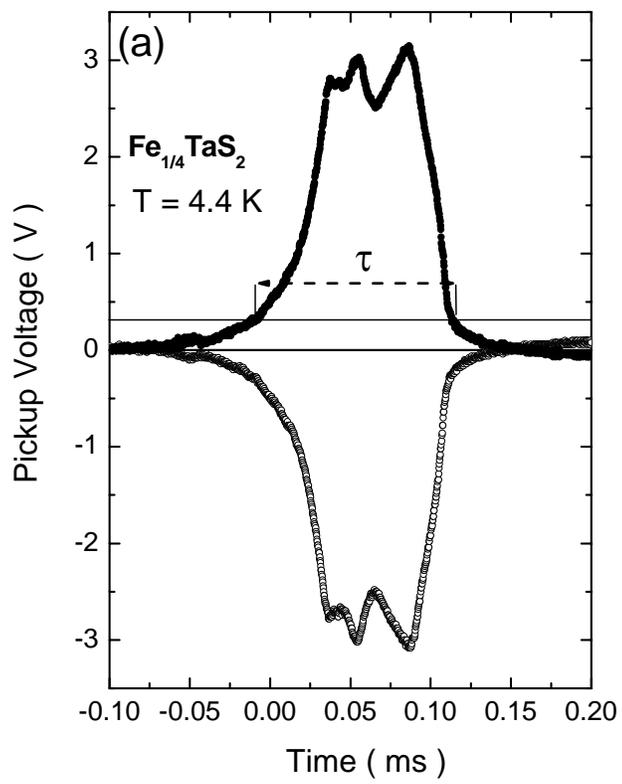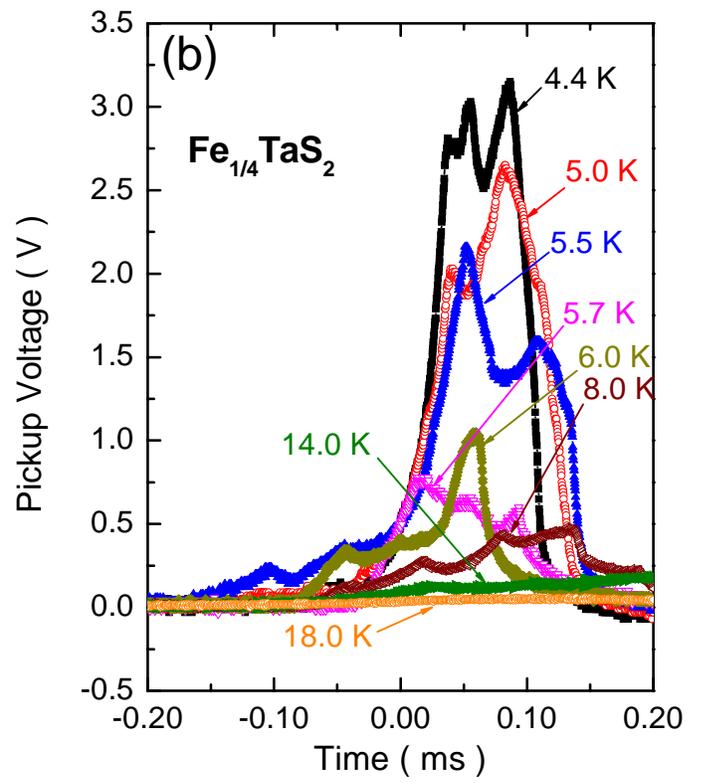

Fig.8

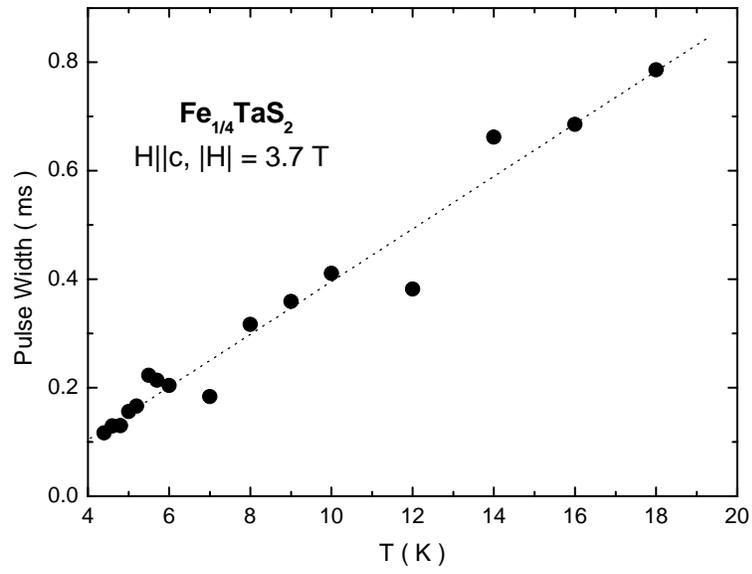

Fig.9

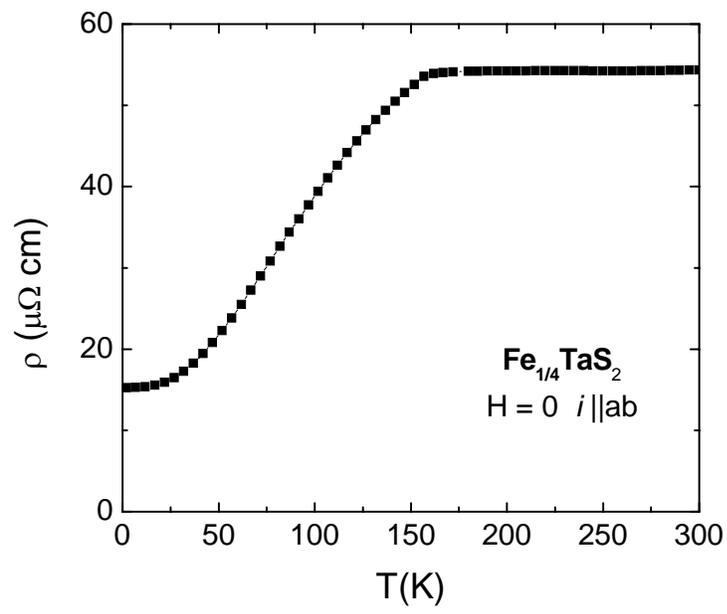

Fig.10

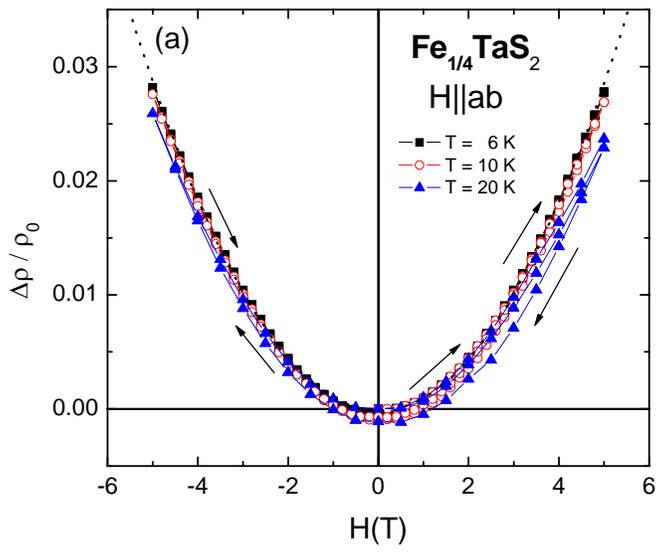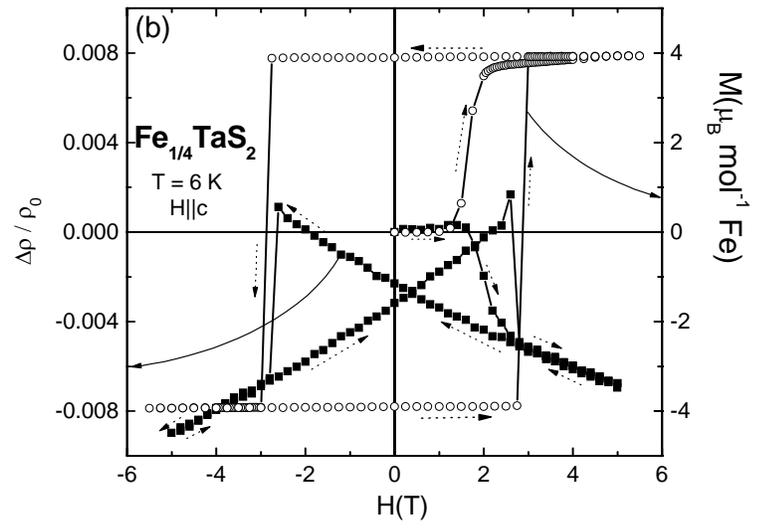